\documentclass[twocolumn,showpacs,preprintnumbers,amsmath,amssymb]{revtex4}

\usepackage{graphicx}
\usepackage{dcolumn}
\usepackage{bm}

\begin{document}


\title{Quantum master equation for a system influencing its environment}

\author{Massimiliano Esposito}
\author{Pierre Gaspard}%
\affiliation{Center for Nonlinear Phenomena and Complex Systems\\
Universit\'e Libre de Bruxelles, Code Postal 231, Campus Plaine, B-1050
Brussels, Belgium.}

\date{\today}

\begin{abstract}
We derive a new perturbative quantum master equation for the
reduced density matrix of a system interacting with an environment
(with a dense spectrum of energy levels). The total system energy (system
plus environment) is constant and finite. This equation takes into
account the finite energy effects of the environment due to the
total energy conservation. This equation is more general than the
common perturbative equations used for describing a system in
interaction with an environment (like the Redfield equation
\cite{Red, GaspRed} or the Cohen-Tannoudji one \cite{Tannoudji})
because these last equations can be deduced from it in the limit
of an infinitely large environment. We apply numerically this
equation to the spin-GORM model. This model represents the
interaction of a two-level system with an environment described by
random matrices. We compare our equation with the exact von
Neumann equation of the total system and show its superiority
compared to the Redfield equation (in the Markovian and 
non-Markovian cases).
\end{abstract}



\maketitle

The study of the quantum dynamics of a system interacting with its
environment is an old but still very important problem in quantum
mechanics and more specifically in non-equilibrium quantum
statistical mechanics. The understanding of such problems is
central for the study of very fundamental physical processes such
as the relaxation to the thermodynamic equilibrium, the decoherence or
the thermalization at the environment temperature.
Actually, one of the more general and of the most used equations to
describe the dynamics of a system interacting with its environment
is the non-Markovian Redfield equation. The other well-known
equations used in the past for the study of such systems (the
Markovian Redfield equation \cite{Red}, the Cohen-Tannoudji
equation \cite{Tannoudji} or the Lindblad equation
\cite{Lindblad}) can be derived from the non-Markovian Redfield
equation \cite{GaspRed}. The Markovian form of this equation was
first derived empirically for understanding NMR experiments
\cite{Red}, but later generalized to the non-Markovian case by
deducing it formally by perturbation theory from the von Neumann
equation of the total system (system plus environment)
\cite{GaspRed}. The fundamental assumption in the Redfield theory
is that the environment is infinite and therefore not affected by
the system. As a consequence of the recent development of
nanotechnology, there is an increasing interest for nanometric systems. 
In nanosystems, one can encounter
situations where purely quantum degrees of freedom 
like the spin are interacting with
degrees of freedom that are dense in energy but with an energy
distribution that varies on small energy scales sometime of the
order of the system energy. In such situations, the Redfield
equation is not valid anymore, the infinite environment
hypothesis fails and the finite energy effects of the total system
have to be taken into account. 
In this paper the plan is as follows: we derive our new equation,
that takes into account the finite energy effects of the total
system, we then apply it to the population dynamics of the
spin-GORM model (a two-level system interacting in a non-diagonal
way with an environment described by Gaussian orthogonal random
matrices) and we finally compare it to the exact von Neumann
equation for the total system and to the Redfield equation. We
will finally conclude presenting the perspectives created by this
work.
The quantum system that we consider has a simple spectrum
(integrable system) given by his Hamiltonian $\hat{H}_{S}$. Its
eigenvalues (respectively its eigenvectors) are given by $\lbrace
E_s \rbrace$ (respectively by $\lbrace \vert s \rangle \rbrace$).
The environment is also a quantum system but with a dense spectrum
(containing many levels forming a quasi-continuum) given by a Hamiltonian
$\hat{H}_{B}$. Its eigenvalues (respectively its eigenvectors) are
given by $\lbrace \epsilon \rbrace$ (respectively by $\lbrace
\vert \epsilon \rangle \rbrace$). The interaction between the
system and the environment is generally taken to be the product of
a system operator $\hat{S}$ and an environment operator $\hat{B}$.
The coupling parameter $\lambda$ measures the intensity of the
interaction between the system and the environment. Therefore, the
Hamiltonian of the total system is given by $\hat{H}_{\rm tot} =
\hat{H}_{S}  + \hat{H}_{B} +  \lambda \hat{S} \hat{B}$. The exact
time evolution of the total system is described by the von Neumann
equation $\dot{\hat{\rho}}(t) = -i\lbrack \hat{H}_{\rm tot} ,
\hat{\rho}(t) \rbrack$, where $\hat{\rho}(t)$ is the density
matrix of the total system. The system dynamics is described by
the reduced density matrix of the system
$\hat{\rho}_S(t)=\textrm{Tr}_{B} \hat{\rho}(t)$. The total system
has a finite constant energy. At initial time, the environment is
in a microcanonical state at energy $\epsilon$.

\begin{figure}[h]
\centering
\rotatebox{0}{\scalebox{0.4}{\includegraphics{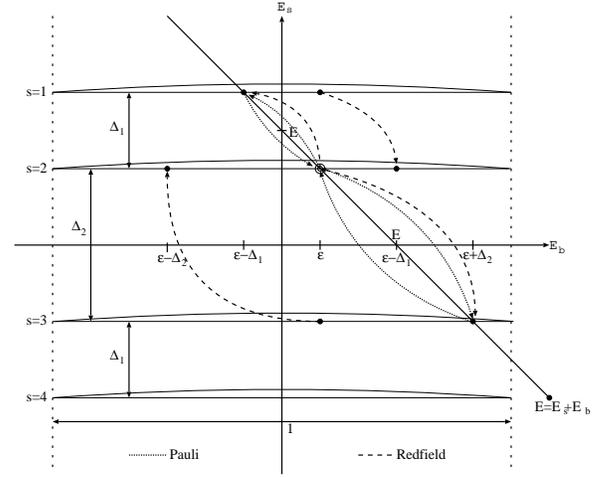}}}
\caption{Representation of the energy exchanges described
respectively by the Cohen-Tannoudji and by the simplified
Markovian version of our new equation for a four level system. The
system energy is represented on the abscissa. The environment
energy (continuum) is represented on the ordinate. The width in
energy of the environment spectrum is fixed to $1$ and its density
of states is supposed semicircular. The total energy of the system
is given by $E$. The initial condition is denoted by two empty
superposed circles. We see that transitions that preserve the
energy of the total system have to occur along the total energy
line crossing the plane. Doing this, they satisfy the Fermi golden
rule for the total system. One can see that only our equation
satisfy this condition. The Cohen-Tannoudji equation describes
transitions that occur along a vertical line at constant
environment energy and is therefore wrong when the system energies
are of the order or larger than the typical energy scale of
variation of the environment density of states. \label{figu1}}
\end{figure}

\begin{figure}[h]
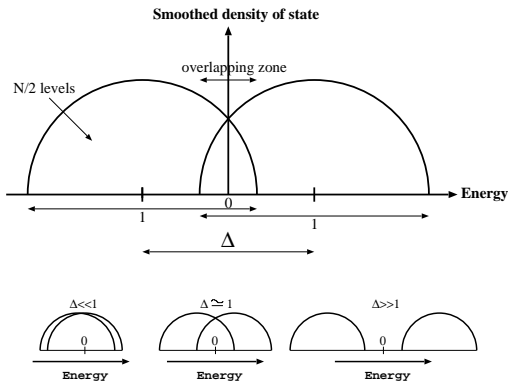

\centerline{\rotatebox{0}{\scalebox{0.6}{\includegraphics{fig2.eps}}}}
\vspace*{0.5cm}
\centerline{\rotatebox{0}{\scalebox{0.5}{\includegraphics{fig3.eps}}}}
\caption{Schematic representation of the smoothed density of
states of the nonperturbed total system ($\lambda=0$) for
different values of $\Delta$.} \label{figu2}
\end{figure}

\begin{figure}[h]
\centerline{\rotatebox{0}{\scalebox{0.8}{\includegraphics{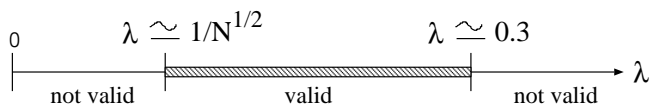}}}}
\caption{Schematic representation of the validity domain of the
kinetic equation in the coupling parameter $\lambda$.} \label{figu3}
\end{figure}

The ansatz in the Redfield derivation is to suppose that the total
density matrix evolves keeping the following form:
\begin{equation}
\hat{\rho}(t)=\hat{\rho}_S(t) \otimes \delta(\epsilon-\hat{H}_B).
\label{formeredfield}
\end{equation}
We see that the environment part of the density matrix does not
evolves, supposing that the environment is not affected by the
dynamics. But as we announced it, we want to include the finite
total energy effects in the dynamics. The main idea (the new
ansatz) is to suppose that the total density matrix can be
described at all times by a density matrix of the following form:
\begin{equation}
\hat{\rho}(t)=\frac{1}{n(\hat{H}_B)} \sum_{s,s'} \vert s \rangle \langle
s' \vert P_{ss'}(\hat{H}_B;t), \label{formegenrhopauli}
\end{equation}
where $n(\epsilon)=\textrm{Tr}_{B} \delta(\epsilon-\hat{H}_B)$ is
the smoothed density of states of the environment. The reduced
density matrix of the system becomes $\hat{\rho}_S(t)= \int
d\epsilon \textrm{Tr}_{B} \delta(\epsilon-\hat{H}_B) \hat{\rho}(t)
= \sum_{s,s'} \vert s \rangle \langle s' \vert \int d\epsilon
P_{ss'}(\epsilon;t)$. The only approximation made by the ansatz
is that we neglect the contribution to the dynamics coming from the
environment coherences. We see that the environment energy can now
depend on the system state. Inserting (\ref{formegenrhopauli}) in
the von Neumann equation, taking the trace over the environment
degrees of freedom and performing a perturbative expansion up to
the second order in $\lambda$, one gets our new equation. For the
population dynamics, this equation takes the following form:
\begin{eqnarray}
\dot{P}_{ss}(\epsilon;t) &=& -2\lambda^2 \sum_{\bar{s},\bar{s}'}
\Big\lbrack  \label{pauligenNM} \\ & &\hspace*{-1.8cm} + \langle s
\vert \hat{S} \vert \bar{s}' \rangle \langle \bar{s}' \vert
\hat{S} \vert \bar{s} \rangle \nonumber \\ & &\hspace*{-1.5cm}
P_{\bar{s}s}(\epsilon;t) \int d\epsilon' F(\epsilon,\epsilon') 
n(\epsilon') \frac{\sin
(E_{\bar{s}}-E_{\bar{s}'}+\epsilon-\epsilon')t}{E_{\bar{s}}-E_{\bar{s}'}+
\epsilon-\epsilon'} \nonumber \\ & &\hspace*{-1.8cm} - \langle s
\vert \hat{S} \vert \bar{s} \rangle  \langle \bar{s}' \vert
\hat{S} \vert s \rangle \nonumber \\ & &\hspace*{-1.5cm}
n(\epsilon) \int d\epsilon' F(\epsilon,\epsilon') 
 P_{\bar{s}\bar{s}'}(\epsilon';t)
\frac{\sin (E_{s}-E_{\bar{s}'}+\epsilon-\epsilon')t}
{E_{s}-E_{\bar{s}'}+\epsilon-\epsilon'} \Big\rbrack \nonumber,
\end{eqnarray}
where $F(\epsilon,\epsilon') =``\vert \langle \epsilon
\vert \hat{B} \vert \epsilon' \rangle \vert^2"$ where the quotes
denote a smoothening over the dense sppectrum
of eigenvalues around $\epsilon$ and $\epsilon'$
\cite{EspoGasppaulgen}.
Performing the same procedure using (\ref{formeredfield}) instead
of (\ref{formegenrhopauli}) gives the non-Markovian Redfield
equation \cite{GaspRed} which can be seen as the particular case of
our equation when the environment density of states varies on a
large energy scale compared to the typical energies of the system.
The Markovian approximation consists in taking the infinite time
limit of the time-dependent coefficients of (\ref{pauligenNM})
using the property $\lim_{\tau \to \infty} \frac{\sin(\xi
\tau)}{\xi}= \pi \delta(\xi)$. Performing this approximation on
our new equation (respectively on the non-Markovian Redfield
equation) gives the Markovian version of our new equation
(respectively the Redfield equation \cite{GaspRed}). If one
further neglects the contributions of the coherences to the
populations evolution and the contributions of the populations to
the coherences evolution, one gets a simplified Markovian version
of our new equation. This equation takes the following form for
the populations dynamics:
\begin{eqnarray}
\dot{P}_{ss}(\epsilon;t) &=& -2\pi \lambda^2 \sum_{s'\ne s} \vert
\langle s \vert \hat{S} \vert s' \rangle \vert^2 
F(\epsilon,E_{s}-E_{s'}+\epsilon)
\nonumber \\ & &\hspace*{2cm} n(E_{s}-E_{s'}+\epsilon)
P_{s,s}(\epsilon;t) \nonumber \\ & & +2\pi \lambda^2 \sum_{s'\ne
s} \vert \langle s \vert \hat{S} \vert s' \rangle \vert^2 
F(\epsilon,E_{s}-E_{s'}+\epsilon) \nonumber \\ & &\hspace*{2cm} n(\epsilon)
P_{s',s'}(E_{s}-E_{s'}+\epsilon;t) . \label{origpaulipop}
\end{eqnarray}
Doing the same for the Redfield equation one get the well known
Cohen-Tannoudji equation \cite{Tannoudji}.
Our simplified Markovian equation
describes the evolution of the total system as a random walk
between the states belonging to the same energy shell with
transition probabilities given by the Fermi golden rule.
This equation may look like the Pauli equation \cite{Pauli}
but it describes the time evolution of
the distributions of populations over the energy of the
environment which is the new feature of our equation. On the other
hand, the difference between our simplified Markovian equation (respectively
the non-Markovian version of our equation) and the Cohen-Tannoudji
equation \cite{Tannoudji} (respectively the non-Markovian Redfield equation) is
the fact that this last equation considers that the
environment density of states is not affected by the system energy.
This is represented in Fig. \ref{figu1}.
We will apply now our equation to the spin-GORM model. This model
describes the evolution of a two-level system that interacts in a
non-diagonal way with a complex environment. Here, the complexity
is supposed to come from many-body interactions like in heavy
nuclei or from a classically chaotic dynamics. Therefore, we
represent all the environment operators by Gaussian orthogonal random
matrices (GORM). The Hamiltonian of the total system is then
$\hat{H}_{\rm tot} = \frac{\Delta}{2} \hat{\sigma}_{z} +
\frac{1}{\sqrt{8N}} \hat{X} + \lambda \hat{\sigma}_x
\frac{1}{\sqrt{8N}} \hat{X}'$, where $\hat{X}$ and $\hat{X}'$ are Gaussian
orthogonal random matrices of size $N$ and 
probability density proportional to $\exp(-\frac{1}{4}{\rm Tr}\hat{X}^2$.
The three fundamental parameters determining the model are $\Delta$,
$\lambda$, and $N$. The smoothed density of states of the
environment is given by the Wigner semicircular $n(\epsilon) =
\frac{4N}{\pi} \sqrt{\frac{1}{4}-\epsilon^2}$
with the convention that $\sqrt{x}=0$ for $x<0$. 
The smoothed density of states of
the nonperturbed total system ($\lambda=0$) is the sum of two
semicircular centered at the two energies $-\frac{\Delta}{2}$
and $\frac{\Delta}{2}$ as depicted in Fig. \ref{figu2}.\\
In this paper, we restrict ourselves to the study of the small
coupling regimes $\lambda \ll 1$ because all the equations
discussed in this paper are obtained perturbatively. Another
restriction is related to the mean level spacing of the
environment. The coupling between the levels (that is of order
$\lambda^2$ because the first order in perturbation theory is zero
due to the non-diagonal nature of the coupling) has to be of the
order or larger than the mean level spacing $\frac{1}{N}$ of the
environment to induce a sufficient interaction between the levels
belonging to the microcanonical energy shell in order to reach a
microcanonical distribution inside these shells. The criterion is
therefore $\lambda^2 \geq \frac{1}{N}$. Of course, the lower bound
disappears in the continuum limit $N \to \infty$. The validity
domain is shown in Fig. \ref{figu3}. The detailed study of
the lower bound of the coupling parameter has been done in
\cite{EspoGaspspingorm}.\\
Applying our equation (\ref{pauligenNM}) to the spin-GORM model in
order to study the system population evolution through
$\hat{\sigma}_z$ (difference between the probability of being in
the upper state of the system minus the probability of being in
the lower one), one gets \cite{EspoGasppaulgen}
\begin{eqnarray}
\langle \hat{\sigma}_z \rangle^{NM}(t) = \int d\epsilon' \left[
P_{++}(\epsilon';t)-P_{--}(\epsilon'+\Delta;t) \right]
\label{zdefpauligen1}
\end{eqnarray}
where
\begin{eqnarray}
\dot{P}_{\pm\pm}(\epsilon;t) &=& \\ & & \hspace*{-1.5cm} -
\frac{\lambda^2}{\pi} P_{\pm\pm}(\epsilon;t) 
\int_{-\frac{1}{2}}^{+\frac{1}{2}} d\epsilon' 
\sqrt{\frac{1}{4}-{\epsilon'}^2} \frac{\sin(\pm
\Delta+\epsilon-\epsilon')t}{(\pm \Delta+\epsilon-\epsilon')}
\nonumber \\ & &\hspace*{-1.5cm} + \frac{\lambda^2}{\pi}
\sqrt{\frac{1}{4}-\epsilon^2} \int_{-\frac{1}{2}}^{+\frac{1}{2}} 
d\epsilon' P_{\mp\mp}(\epsilon';t)
\frac{\sin(\pm \Delta + \epsilon - \epsilon') t} {(\pm \Delta +
\epsilon - \epsilon')}. \nonumber \label{paulispingoeNM11}
\end{eqnarray}
An important remark is that the Markovian and the simplified
Markovian version of our equation are the same for the spin-GORM
model. Neglecting the coherences contributions to the
populations dynamics is not necessary here. This has also the
consequence that the Markovian Redfield equation and the
Cohen-Tannoudji are the same.

\begin{figure}[h]
\centerline{\rotatebox{0}{\scalebox{0.25}{\includegraphics{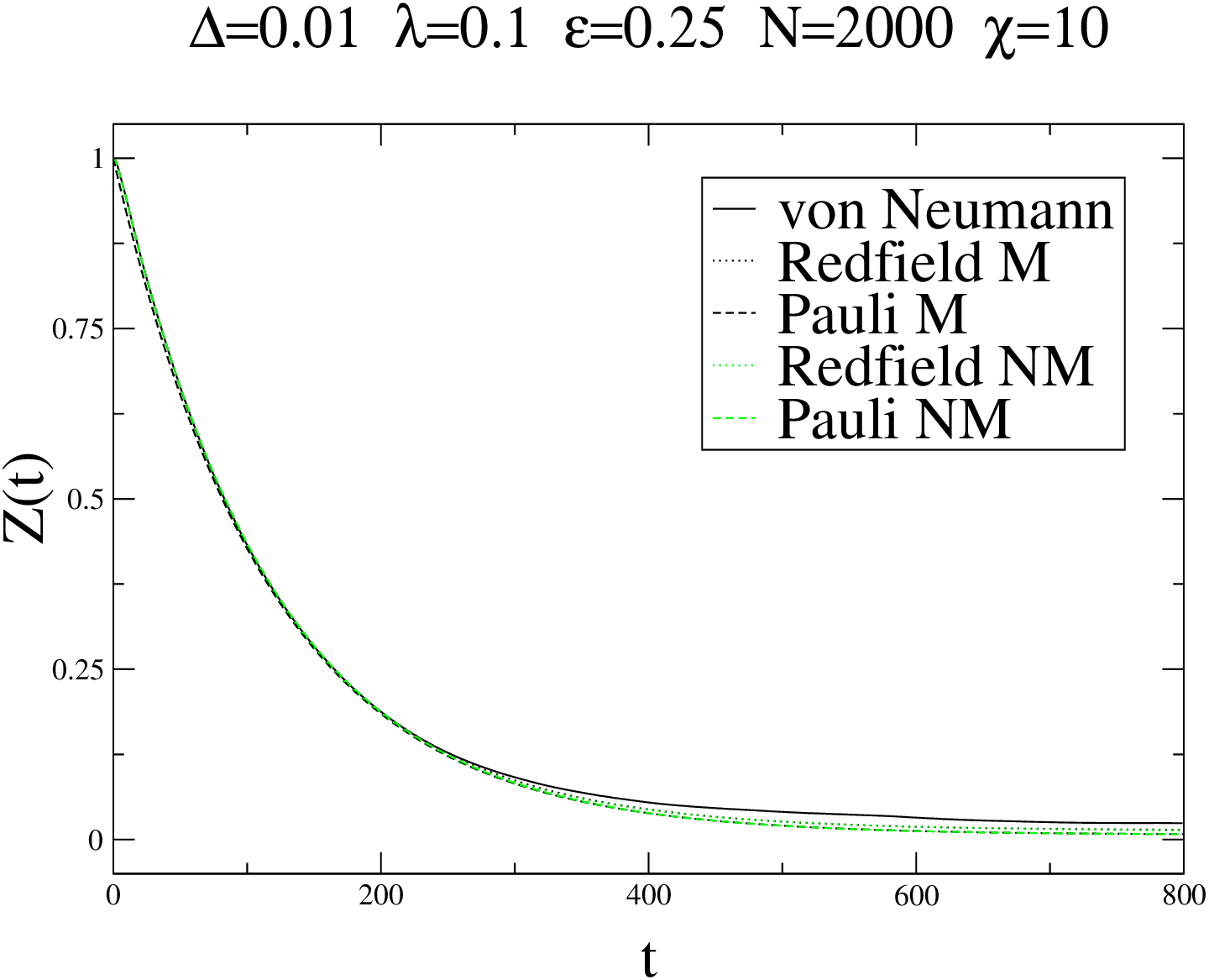}}}}
\vspace*{0.5cm}
\centerline{\rotatebox{0}{\scalebox{0.25}{\includegraphics{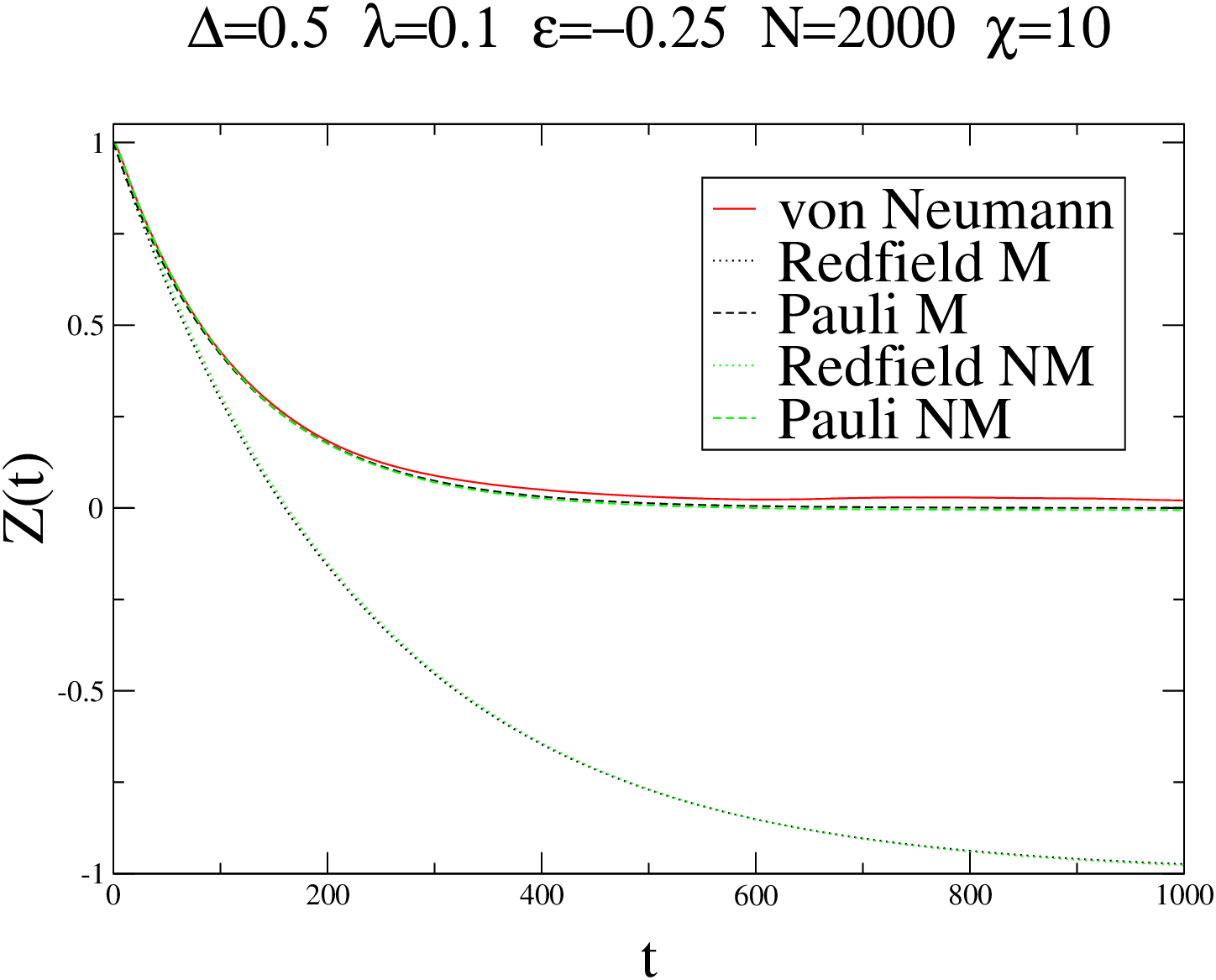}}}}
\vspace*{0.5cm}
\centerline{\rotatebox{0}{\scalebox{0.25}{\includegraphics{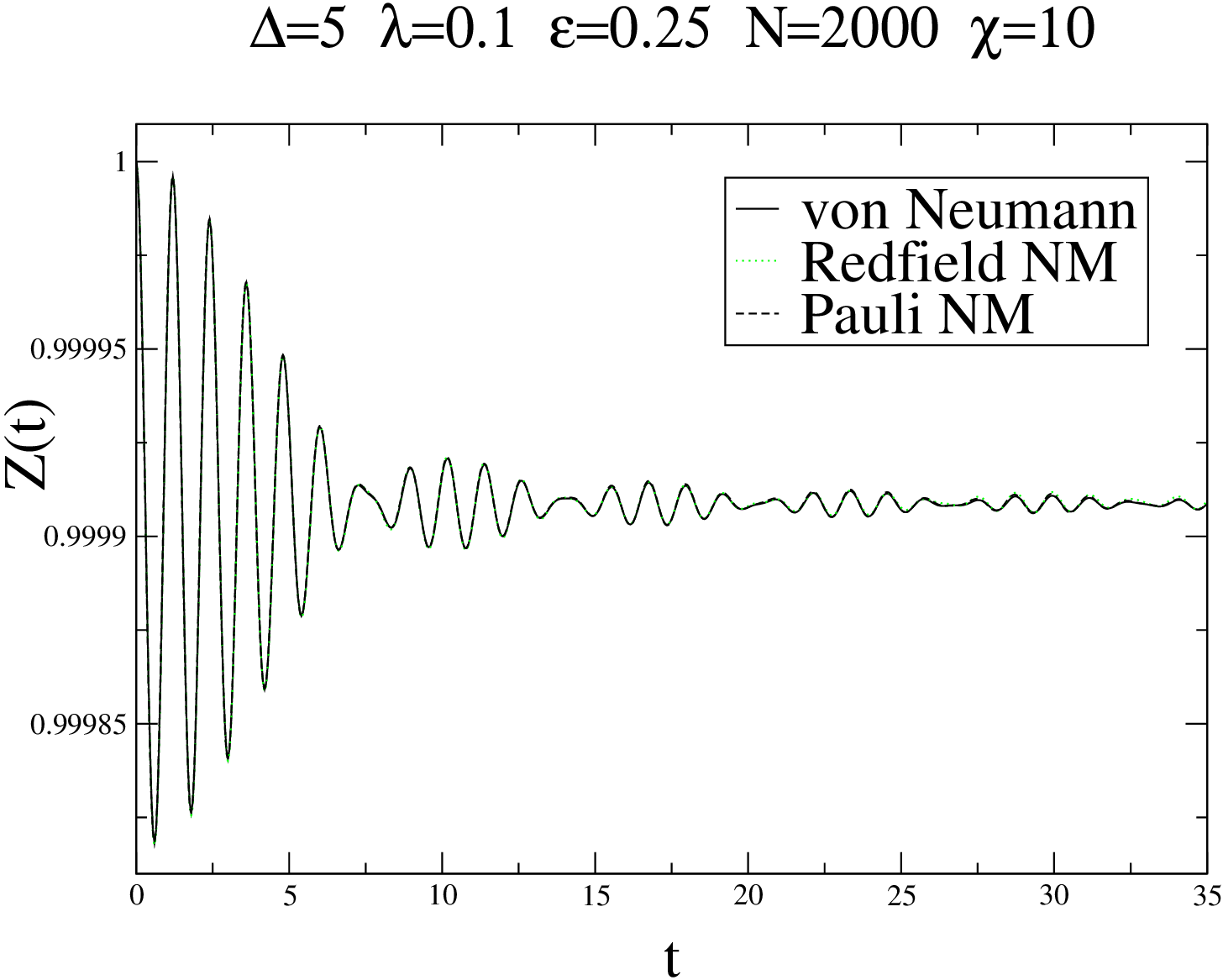}}}}
\caption{Two-level system dynamics for the spin-GORM model.}
\label{figu4}
\end{figure}

The Markovian version of our equation (\ref{zdefpauligen1})
becomes the following equation for the spin-GORM model: $\langle
\hat{\sigma}_z \rangle^{M}(t) =
P_{++}(\epsilon;t)-P_{--}(\epsilon+\Delta;t)$. In this case, we
get an analytical solution describing an exponential decay to
equilibrium:
\begin{eqnarray}
\langle \hat{\sigma}_z \rangle^{M}(t)= \left[ \langle
\hat{\sigma}_z \rangle^{M}(0)-\langle \hat{\sigma}_z
\rangle^{M}_{\infty} \right] e^{- \gamma t} + \langle
\hat{\sigma}_z \rangle^{M}_{\infty} , \label{paulispingoeMZt}
\end{eqnarray}
where the relaxation rate is given by
\begin{equation}
\gamma= \lambda^2 \left(\sqrt{\frac{1}{4}-(\epsilon)^2}
+\sqrt{\frac{1}{4}-(\epsilon+\Delta)^2}\right) \nonumber,
\label{paulispingoeMrate}
\end{equation}
and the equilibrium population by
\begin{equation}
\langle \hat{\sigma}_z \rangle^{M}_{\infty}
=\frac{\sqrt{\frac{1}{4}-(\epsilon)^2}-
\sqrt{\frac{1}{4}-(\epsilon+\Delta)^2}}
{\sqrt{\frac{1}{4}-(\epsilon)^2} +
\sqrt{\frac{1}{4}-(\epsilon+\Delta)^2}} \nonumber .
\label{paulispingoeMZinfini}
\end{equation}
Notice that the relaxation rate and the equilibrium population are
independent of the initial condition.\\
We now start the discussion of the comparison between the
different equations based on the numerical simulations of
Fig. \ref{figu4}. We always take $\langle \hat{\sigma}_z
\rangle^{M}(0)=1$ in our numerical simulations. The initial state
of the environment is characterized by the energy of the
microcanonical distribution $\epsilon$. The width of the energy
shell is always $\delta \epsilon = 0.05$. The curves are
averages over $\chi=10$ realizations of GORM of size
$N=2000$ of the Hamiltonian, and over the different eigenstates of
the environment energy shell.\\
As we announced it, our equation is necessary when the system
energy $\Delta$ is of the order of magnitude of the energy scale of
variation of typical environment smooth density of states:
$n(\epsilon+\Delta) \neq n(\epsilon)$. But when $\Delta$ is small
enough and therefore $n(\epsilon+\Delta) \approx n(\epsilon)$, our
equation reduces to the non-Markovian Redfield equation. This can
be well seen comparing Figs. \ref{figu4}A and B. 
In Fig. \ref{figu4}A, our Markovian (M) and
non-Markovian (NM) equations are very close to the Redfield 
M and NM ones, but this is not the
case anymore in Fig. \ref{figu4}B. 
Another important point is the validity of
the Markovian approximation. This approximation has the effect of
constraining the dynamics inside the microcanonical energy shell of the
non-perturbed spectrum according to the Fermi golden rule. At short
time, because $\frac{\sin(\pm \Delta+\epsilon-\epsilon')t}{(\pm
\Delta+\epsilon-\epsilon')}$ is not yet a delta function, the
non-Markovian equation has the possibility of describing a spread
of the probability in energy around the microcanonical energy shell. When
the delta contribution to the dynamics is important, the
non-Markovian effect are very small, as in Figs. \ref{figu4}A and B, and can
only be seen on very short time scales. But when this contribution
is small or zero, like in Fig. \ref{figu4}C (because the microcanonical energy shell
is not inside the overlapping zone as seen in Fig. \ref{figu2}), then
the non-Markovian effects become very important. In Fig. \ref{figu4}C, the
Markovian curves are not represented because they completely
miss the dynamics (they predict no evolution $\langle \hat{\sigma}_z
\rangle=1$). Fig. \ref{figu4}C represents therefore a pure non-Markovian
dynamics to which only the non-central part of $\frac{\sin(\pm
\Delta+\epsilon-\epsilon')t} {(\pm \Delta+\epsilon-\epsilon')}$
contributes.\\
We can conclude saying that our new equation is always valid in
the small coupling limit independently of the energy ratio between
the system and the environment. It is therefore an important
equation for the study of nanosystems. This equation reduces
to the Redfield equation for very small system energies compared
to the energy variations of the environment density of states. It can
be shown that it is also in this limit that the equilibrium
values of the system populations thermalize to a canonical
distribution corresponding to the microcanonical temperature of the
environment \cite{EspoGaspspingorm}. The coherences
dynamics will be investigated in future work.\\
The authors thank Professor G. Nicolis for support and
encouragement in this research, as well as D. Cohen for several
very fruitful discussions during his visit to Brussels. M. E. is
supported by the Fond pour la formation \`{a} la Recherche dans
l'Industrie et dans l'Agriculture, and P.G. by the National Fund
for Scientific Research (F.~N.~R.~S. Belgium).

\end{document}